\documentclass[aps,prb,twocolumn,groupedaddress]{revtex4-1}

\usepackage{graphicx}
\usepackage{epsfig}
\usepackage{dcolumn}
\usepackage{bm}
\usepackage{tabularx}
\usepackage{hyperref}
\usepackage{xcolor}
\usepackage{amsmath}
\usepackage{subfigure}
\usepackage{float}
\usepackage{booktabs,dcolumn}
\usepackage{caption}
\usepackage{textcomp}
\usepackage{amssymb}

\hypersetup{citecolor=black, filecolor= black, linkcolor= black, urlcolor= black}

\begin{document}

\title{High-Temperature Superconductivity in the Ti--H System at High Pressures}
\author{Jin Zhang}
\affiliation{Department of Physics and Astronomy, Washington State University, Pullman, WA 99164, USA}

\author{Jeffrey M. McMahon}
\email{jeffrey.mcmahon@wsu.edu}
\affiliation{Department of Physics and Astronomy, Washington State University, Pullman, WA 99164, USA}

\author{Artem R. Oganov}
\email{A.Oganov@skoltech.ru}
\affiliation{Skolkovo Institute of Science and Technology, Skolkovo Innovation Center, 5 Nobel St., Moscow, 143026, Russia}
\affiliation{Moscow Institute of Physics and Technology, 9 Institutskiy Lane, Dolgoprudny 141700, Russia}
\affiliation{International Center for Materials Discovery, Northwestern Polytechnical University, Xi'an, Shaanxi 710072, PR China}

\author{Xinfeng Li}
\affiliation{Sino-French Institute of Nuclear Engineering and Technology, Sun Yat-sen University, Zhuhai, Guangzhou 519082, China}

\author{Xiao Dong}
\affiliation{Key Laboratory of Weak-Light Nonlinear Photonics and School of Physics, Nankai University, Tianjin 300071, China}

\author{Huafeng Dong}
\affiliation{School of Physics and Optoelectronic Engineering, Guangdong University of Technology, Guangzhou 510006, China}

\author{Shengnan Wang}
\affiliation{Department of Geosciences, Center for Materials by Design, and Institute for Advanced Computational Science, State University of New York, Stony Brook, NY 11794-2100, USA}

\begin{abstract}
Search for stable high-pressure compounds in the Ti--H system reveals the existence of titanium hydrides with new stoichiometries, including \emph{Ibam}-Ti$_2$H$_5$, \emph{I}4/\emph{m}-Ti$_5$H$_{13}$, \emph{I}$\bar{4}$-Ti$_5$H$_{14}$, \emph{Fddd}-TiH$_4$, \emph{Immm}-Ti$_2$H$_{13}$, \emph{P}$\bar{1}$-TiH$_{12}$, and \emph{C}2/\emph{m}-TiH$_{22}$. Our calculations predict \emph{I}4/\emph{mmm} $\rightarrow$ \emph{R}$\bar{3}$\emph{m} and \emph{I}4/\emph{mmm} $\rightarrow$ \emph{Cmma} transitions in TiH and TiH$_2$, respectively. Phonons and the electron--phonon coupling of all searched titanium hydrides are analyzed at high pressure. It is found that \emph{Immm}-Ti$_2$H$_{13}$ rather than the highest hydrogen content \emph{C}2/\emph{m}-TiH$_{22}$, exhibits the highest superconducting critical temperature \emph{T}$_{c}$. The estimated \emph{T}$_{c}$ of \emph{Immm}-Ti$_2$H$_{13}$ and \emph{C}2/\emph{m}-TiH$_{22}$ are respectively 127.4--149.4 K ($\mu^{*}$=0.1-0.15) at 350 GPa and 91.3--110.2 K at 250 GPa by numerically solving the Eliashberg equations. One of the effects of pressure on \emph{T}$_{c}$ can be attributed to the softening and hardening of phonons with increasing pressure.
\end{abstract}
\maketitle

\section{Introduction}

Enthusiasm for discovering high-temperature superconductors never cease \cite{ginzburg1999problems}, since solid mercury was discovered to have zero electrical resistance below 4.2 K in 1911\cite{van2010discovery}. In recent years, especially at high pressure, the record of the critical temperature (\emph{T}$_c$) of superconductivity has been quickly and repeated to be broken in both experimental and theoretical studies, rendering the ultimate goal for synthesizing a room-temperature superconductor (\emph{T}$_c$ at around 298 K) appears to be within reach. In 2014, first-principles calculation\cite{duan2014pressure} based on density functional theory (DFT) predicted the \emph{T}$_c$ of \emph{Im}$\bar{3}$\emph{m}-H$_3$S to be around 191 K--204 K at 200 GPa. Subsequently, diamond anvil cell (DAC) experiment in 2015\cite{drozdov2015conventional} verified this prediction and reported the \emph{T}$_c$ of sulfur hydride of 203 K by compressing hydrogen sulfide to 150 GPa. In 2017, DFT calculation\cite{liu2017potential} estimated \emph{T}$_c$ of \emph{Fm}$\bar{3}$\emph{m}-LaH$_{10}$ to be 274--286 K at 210 GPa and of \emph{Fm}$\bar{3}$\emph{m}-YH$_{10}$ to be 305--326 K (the highest theoretically-calculated \emph{T}$_c$ for simple binary systems so far\cite{bi2018search}) at 250 GPa. Soon afterwards, the teams of Hemley\cite{somayazulu2019evidence} and Eremets\cite{drozdov2019superconductivity} observed lanthanum hydride (\emph{Fm}$\bar{3}$\emph{m}-LaH$_{10}$) superconducting under the pressure (170-200 GPa) at around 250--260 K, which is the highest \emph{T}$_c$ that has been experimentally confirmed.  Although the effect of pressure on superconductivity is not fully understood\cite{lorenz2005high,gao1994superconductivity}, these new record high-\emph{T}$_c$ superconductors are conventional, phonon-mediated ones. Based on Bardeen--Cooper--Schrieffer (BCS) or Migdal--Eliashberg theories, the pressure affects the \emph{T}$_c$ by making impact on the electronic and phonon parameters, e.g. electronic density of states at the Fermi level, average phonon frequency, and electron-phonon coupling (EPC) constant.

Motivation for investigating superconductivity of hydrides under pressure originally came from both the possibility that metallic hydrogen under high pressure could be a high-temperature superconductor\cite{ashcroft1968metallic} and from the viewpoint that the pressure of metallization of hydrogen-rich solids can be considerably lower than that of pure hydrogen\cite{ashcroft2004hydrogen,ashcroft2004bridgman}. Since carrying out the high-pressure experiments is expensive and technically challenging, many of the investigations on these superconductors are performed using calculations and crystal structure prediction techniques. Besides \emph{Im}$\bar{3}$\emph{m}-H$_3$S, \emph{Fm}$\bar{3}$\emph{m}-LaH$_{10}$ and \emph{Fm}$\bar{3}$\emph{m}-YH$_{10}$ (mentioned above), the calculated \emph{T}$_c$ of some predicted structures are as follows: \emph{R}$\bar{3}$\emph{m}-LiH$_6$ is 82 K at 300 GPa\cite{xie2014superconductivity}, \emph{Im}$\bar{3}$\emph{m}-MgH$_6$ is 271 K at 400 GPa\cite{feng2015compressed}, \emph{Im}$\bar{3}$\emph{m}-CaH$_6$ is 220--235 K at 150 GPa\cite{wang2012superconductive}, \emph{I}4$_1$\emph{md}-ScH$_9$ is 233 K at 300 GPa\cite{abe2017hydrogen}, \emph{Cmcm}-ZrH is 11 K at 120 GPa\cite{li2017phase}, \emph{P}2$_1$/\emph{m}-HfH$_2$ is 11--13 K at 260 GPa\cite{liu2015first}, \emph{Fdd}2-TaH$_6$ is 124--136 K at 300 GPa\cite{zhuang2017pressure}, \emph{Pm}$\bar{3}$\emph{m}-GeH$_3$ is 140 K at 180 GPa\cite{gao2010high}, \emph{P}6/\emph{mmm}-LaH$_{16}$ is 156 K at 200 GPa\cite{kruglov2018superconductivity}, \emph{C}2/\emph{m}-SnH$_{14}$ is 86--97 K at 300 GPa\cite{esfahani2016superconductivity}, \emph{Im}$\bar{3}$\emph{m}-H$_3$Se is 131 at 200 GPa\cite{flores2016high}, \emph{P}6/\emph{mmm}-H$_4$Te is 95--104 at 170 GPa\cite{zhong2016tellurium}. Almost all binary hydrides systems have been computationally studied by now, at least crudely, see an overview in Ref. \onlinecite{semenok2018distribution}.

Transition metal hydrides can form a variety of stable stoichiometries and have lower metallization pressure compared with other hydrides. Especially, those with high hydrogen content often contain unexpected hydrogen groups and exhibit intriguing properties. Titanium is such transition metal that inspires us to study the titanium hydrides under high pressure. At ambient conditions, TiH$_2$ crystallizes in a tetragonal structure (\emph{I}4/\emph{mmm}), which transforms into a cubic phase (\emph{Fm}$\bar{3}$\emph{m}) at temperature increasing to 310 K\cite{yakel1958thermocrystallography,san1987h}. DAC experiments\cite{vennila2008phase,kalita2010equation,endo2013phase} indicated that \emph{I}4/\emph{mmm}-TiH$_2$ remains stable at the pressure up to 90 GPa at ambient temperature. The theoretically estimated \emph{T}$_c$ is 6.7 K ($\lambda$=0.84, $\mu^{*}$=0.1) for \emph{Fm}$\bar{3}$\emph{m}-TiH$_2$ and 2 mK ($\lambda$=0.22, $\mu^{*}$=0.1) for \emph{I}4/\emph{mmm}-TiH$_2$\cite{shanavas2016electronic} at ambient pressure.

In this paper, the crystal structures and superconductivity of titanium hydrides at pressures up to 350 GPa are systematically studied. In addition to \emph{I}4/\emph{mmm}-TiH$_2$, several new stoichiometries and phases are found at high pressure by a first-principles evolutionary algorithm. The predicted TiH$_{22}$ becomes thermodynamically stable at pressure above 235 GPa and contains H$_{20}$ units in its crystal structure. The dynamical stability of all high-pressure phases was verified by calculations of phonons throughout the Brillouin zone. Three different approaches are utilized to determine the superconducting \emph{T}$_c$. The predicted \emph{T}$_c$ (numerical solution from the Eliashberg equations) for \emph{C}2/\emph{m}-TiH$_{22}$ and \emph{Immm}-Ti$_2$H$_{13}$ are 91.3--110.2 K (at 250 GPa) and 127.4--149.4 K (at 350 GPa), respectively. Our work provides clear guidance for future experimental investigation of potential high-temperature superconductivity in titanium hydrides under pressure.

\section{Computational methodology}

Variable-compositional prediction of stable compounds in the Ti--H system was performed at 0, 50, 100, 150, 200, 250, 300, and 350 GPa through first-principles evolutionary algorithm (EA), as implemented in the USPEX code\cite{oganov2006crystal,oganov2011evolutionary,lyakhov2013new}. Structure relaxations were based on DFT within the Perdew--Burke--Ernzerhof (PBE) generalized gradient approximation (GGA) exchange--correlation functional\cite{perdew1996generalized}, as implemented in the VASP package\cite{kresse1996efficient}. The initial generation consisting of 120 structures was created using random symmetric generator. Structures in the following generations were created from the previous generation using heredity (40\%), lattice mutation (20\%), random symmetric generator (20\%) and transmutation (20\%) operators. The electron--ion interaction was described by projector-augmented wave (PAW) potentials\cite{blochl1994projector,kresse1999ultrasoft}, with 3\emph{p}$^6$4\emph{s}$^2$3\emph{d}$^4$ and 1\emph{s}$^2$ shells treated as valence for Ti and H, respectively. Structures predicted to be stable or low-enthalpy metastable were then carefully reoptimized to construct convex hull and phase diagram at each pressure. Brillouin zone (BZ) was sampled using $\Gamma$-centered uniform \emph{k}-meshes (2$\pi\times0.05$ {\AA}$^{-1}$) and the kinetic energy cutoff for the plane-wave basis set was 600 eV.

Phonon calculations were carried out using the finite-displacement method as implemented in the Phonopy \cite{togo2008first} codes, using VASP to calculate the force constants matrix, as well as density functional perturbation theory (DFPT)\cite{baroni2001phonons} in the {\sc Quantum ESPRESSO} ({\sc QE}) package\cite{giannozzi2009quantum,QE-2017}. Results of these two methods were in perfect agreement. The EPC coefficients were calculated using DFPT in {\sc QE}, the norm-conserving pseudopotentials (tested by comparing the phonon spectra with the results calculated from Phonopy codes) and the PBE functional were used. Convergence tests show that 120 Ry is a suitable cutoff energy for the plane wave basis set in the {\sc QE} calculation. A 4$\times$4$\times$4 \emph{q}-mesh was used in the phonon and electron--phonon calculations.

\emph{T}$_{c}$ is estimated using three approaches: by numerically solving Eliashberg equations, and solving approximate Allen--Dynes formula, and the (latter-)modified McMillan formula. Starting from BCS theory, several first-principles Green's function methods had been proposed to calculate the superconducting properties. Migdal--Eliashberg formalism is one of these, and can accurately describe conventional superconductors\cite{grimaldi1999physical}. Within the Migdal approximation\cite{migdal1958interaction}, the adiabatic ratio $\lambda$$\omega$$_{D}$/$\epsilon$$_{F}$ ($\simeq$$\sqrt{m^{*}/M}$) is small, since the vertex correction O($\sqrt{m^{*}/M}$) can compare to the bare vertex and then be neglected. In the adiabatic ratio, \emph{m}$^{*}$ is the electron effective mass, \emph{M} is the ion mass, $\omega$$_{D}$ is Debye frequency and $\epsilon$$_{F}$ is Fermi energy. Then, \emph{T}$_{c}$ can be calculated by solving two nonlinear Eliashberg equations (or isotropic gap equations) for the Matsubara gap (or superconducting order parameter) $\Delta$$_n$$\equiv$$\Delta$(\emph{i}$\omega$$_i$) and electron mass renormalization function (or wavefunction renormalization factor) \emph{Z}$_n$$\equiv$\emph{Z}(\emph{i}$\omega$$_i$) along the imaginary frequency axis (\emph{i}=$\sqrt{-1}$),
\begin{equation}\label{eq0}
{\textstyle \Delta_{n}Z_{n}=\frac{\pi}{\beta }\sum_{m=-M}^{M}\frac{\lambda (\omega_{n}-\omega_{m})-\mu ^{*}\theta (\omega_{c}-|\omega_{m}|))}{\sqrt{\omega_{m}^{2}+\Delta_{m}^{2}}}\Delta_{m}}
\end{equation}
and
\begin{equation}\label{eq1}
{\textstyle Z_{n}=1+\frac{\pi}{\beta \omega _{n}}\sum_{m=-M}^{M}\frac{\lambda (\omega _{n}-\omega _{m})}{\sqrt{\omega _{m}^{2}+\Delta _{m}^{2}}}\omega_{m}},
\end{equation}
where $\beta$=1/\emph{k}$_{B}$\emph{T}, \emph{k}$_{B}$ is the Boltzmann constant, $\mu^{*}$ denotes the Coulomb pseudopotential, $\theta$ is the Heaviside function, $\omega_{c}$ is the phonon cut off frequency: $\omega_{c}$=3$\omega_{\textit{max}}$, $\omega_{max}$ is the maximum phonon frequency, $\omega_{n}$=($\pi/\beta$)(2\emph{n}-1) is the \emph{n}th fermion Matsubara frequency with n=0,$\pm$1,$\pm$2,..., the pairing kernel for electron--phonon interaction possesses the form $\lambda(\omega _{n}-\omega_{m} )=2\int_{0}^{\omega_{\textit{max}} }\frac{\alpha ^{2}F(\omega )\omega }{\omega ^{2}+(\omega _{n}-\omega _{m})^{2}}d\omega$ and $\alpha ^{2}F(\omega )$ represents the Eliashberg spectral function. A derivation of isotropic Eliashberg gap equations was given in detail by Allen and Mitrovi{\'c}\cite{allen1983theory}. The important feature of the gap equations is that all the involved quantities only depend on the normal state, and then can be calculated first principles. At each temperature \emph{T}, the coupled equations need to be solved iteratively until self-consistency. \emph{T}$_{c}$ is defined as the temperature at which the Matsubara gap $\Delta$$_n$ becomes zero. The Eliashberg equations have been solved numerically for 2201 Matsubara frequencies (\emph{M}=1100), in this paper. A detailed discussion of this numerical method was presented in Refs. \onlinecite{szczesniak2006numerical,szcze2012superconducting}.

In addition to the above numerical method, \emph{T}$_{c}$ can also be obtained by other two analytical formulas. The first one was introduced by Allen and Dynes and the second one was initially proposed by McMillan and later modified by them. The former formula is given as:
\begin{equation}\label{eq2}
{\textstyle T_{c}=f_{1}f_{2}\frac{\omega_{\textit{log}}}{1.20}\textit{exp}\left(-\frac{1.04(1+\lambda ))}{\lambda-\mu^{*}(1+0.62\lambda)}\right)},
\end{equation}
where the logarithmic average frequency is defined as $\omega_{\textit{log}}=\textit{exp}\left ( \frac{2}{\lambda }\int_{0}^{\omega _{\textit{max}}} d\omega\frac{\alpha^{2}F(\omega ) }{\omega } \textit{ln}(\omega )\right )$, the isotropic EPC constant, which is a dimensionless measure of the average strength of the EPC, can be defined as: $\lambda =2\int_{0}^{\omega _{\textit{max}}}d\omega\frac{\alpha ^{2}F(\omega )}{\omega }$, and \emph{f}$_{1}$ and \emph{f}$_{2}$ are strong coupling correction and shape correction, respectively. These two factors are
\begin{equation}\label{eq3}
{\textstyle f_{1}=\left \{ 1+\left [ \frac{\lambda }{2.46(1+3.8\mu ^{*}))} \right ]^{\frac{3}{2}} \right \}^{\frac{1}{3}}}
\end{equation}
and
\begin{equation}\label{eq4}
{\textstyle f_{2}=1+\frac{\left ( \frac{\bar{\omega}_{2} }{\omega_{\textit{log}} } -1\right )\lambda^{2}}{\lambda^{2}+3.312(1+6.3\mu^{*})^{2}\left ( \frac{\bar{\omega}_{2}}{\omega_{\textit{log}} } \right )^{2}}},
\end{equation}
here, $\bar{\omega}_{2}$ is defines as: $\bar{\omega}_{2} =\left ( \frac{2}{\lambda }\int_{0}^{\omega_{\textit{max}}} d\omega  \alpha ^{2}F(\omega )\omega \right )^{\frac{1}{2}}$. Generally, when $\lambda$ is small, the correction factors \emph{f}$_{1}$ and \emph{f}$_{2}$ are negligible. Therefore, requiring \emph{f}$_{1}$\emph{f}$_{2}$ to be 1 gives the Allen--Dynes modified McMillan equation (only changing a prefactor from $\frac{\Theta_{D}}{1.45}$ in McMillan equation to $\frac{\omega_{\textit{log}}}{1.2}$),
\begin{equation}\label{eq5}
{\textstyle T_{c}=\frac{\omega_{\textit{log}}}{1.20}\textit{exp}\left ( -\frac{1.04(1+\lambda ))}{\lambda-\mu^{*}(1+0.62\lambda )} \right )}.
\end{equation}
Regardless of which of the above three methods is used to calculate \emph{T}$_{c}$, two main input quantities are needed. One is the Coulomb pseudopotential $\mu^{*}$, which models the depairing interaction between the electrons. However, $\mu^{*}$ is hard to calculate from first principles. Herein, we used standard values $\mu^{*}$=0.1 and 0.15. Another one is the Eliashberg spectral function $\alpha ^{2}F(\omega )$, which models the coupling of phonons to electrons on the Fermi surface. $\alpha ^{2}F(\omega )$ can be calculated as\cite{chan2012electron}
\begin{equation}\label{eq6}
{\textstyle \alpha ^{2}F(\omega )=\frac{1}{2\pi N(\epsilon _{F})}\sum_{\mathbf{q}\nu }\delta (\omega -\omega _{\mathbf{q}\nu })\frac{\gamma _{\mathbf{q}\nu }}{\omega _{\mathbf{q}\nu }}},
\end{equation}
where \emph{N}($\epsilon_{F}$) is the density of states at the Fermi level per unit cell per spin, $\mathbf{q}$ is the wavevector, $\omega_{\mathbf{q}}$ is the $\mathbf{q}$-point weight, $\omega _{\mathbf{q}\nu }$ is the screened phonon frequency, and  $\gamma _{\mathbf{q}\nu }$ is the phonon linewidth, which is determined exclusively by the electron--phonon matrix elements $g_{mn}^{\nu }(\mathbf{k},\mathbf{q})$ with states on the Fermi surface, is given as:
\begin{equation}\label{eq7}
\begin{split}
\text{$\gamma_{\mathbf{q}\nu}$} =
& \text{$\pi\omega_{\mathbf{q}\nu}\sum_{mn}\sum_{\mathbf{k}}\omega_{\mathbf{k}}\left | g_{mn}^{\nu }(\mathbf{k},\mathbf{q}) \right |^{2}\delta (\epsilon_{m,\mathbf{k}+\mathbf{q}}-\epsilon_{F})$} \\
& \text{$\times \delta (\epsilon_{n,\mathbf{k}}-\epsilon_{F})$},
\end{split}
\end{equation}
where $\omega_{\mathbf{k}}$ is the $\mathbf{k}$-point weight normalized to 2 in order to account for the spin degeneracy in spin-unpolarized calculations. $g_{mn}^{\nu }(\mathbf{k},\mathbf{q})$ is described as:
\begin{equation}\label{eq8}
{\textstyle g_{mn}^{\nu }(\mathbf{k},\mathbf{q})=\left ( \frac{\hbar}{2M\omega _{\mathbf{q}\nu }} \right )^{1/2} \left \langle m,\mathbf{k}+\mathbf{q}|\delta _{\mathbf{q}\nu }V_{SCF}|n,\mathbf{k} \right \rangle},
\end{equation}
here, $|n,k\rangle$ is the bare electronic Bloch state, \emph{M} is the ionic mass, and $\delta _{\mathbf{q}\nu }V_{SCF}$ is the derivative of the self-consistent potential with respect to the collective ionic displacement corresponding to the phonon wavevector $\mathbf{q}$ and mode $\nu$. In this work, $g_{mn}^{\nu }(\mathbf{k},\mathbf{q})$ is calculated within the harmonic approximation, using {\sc QE} package.

\section{Results and discussions}

\begin{center}
\begin{figure*}
\includegraphics[angle=0,width=1\linewidth]{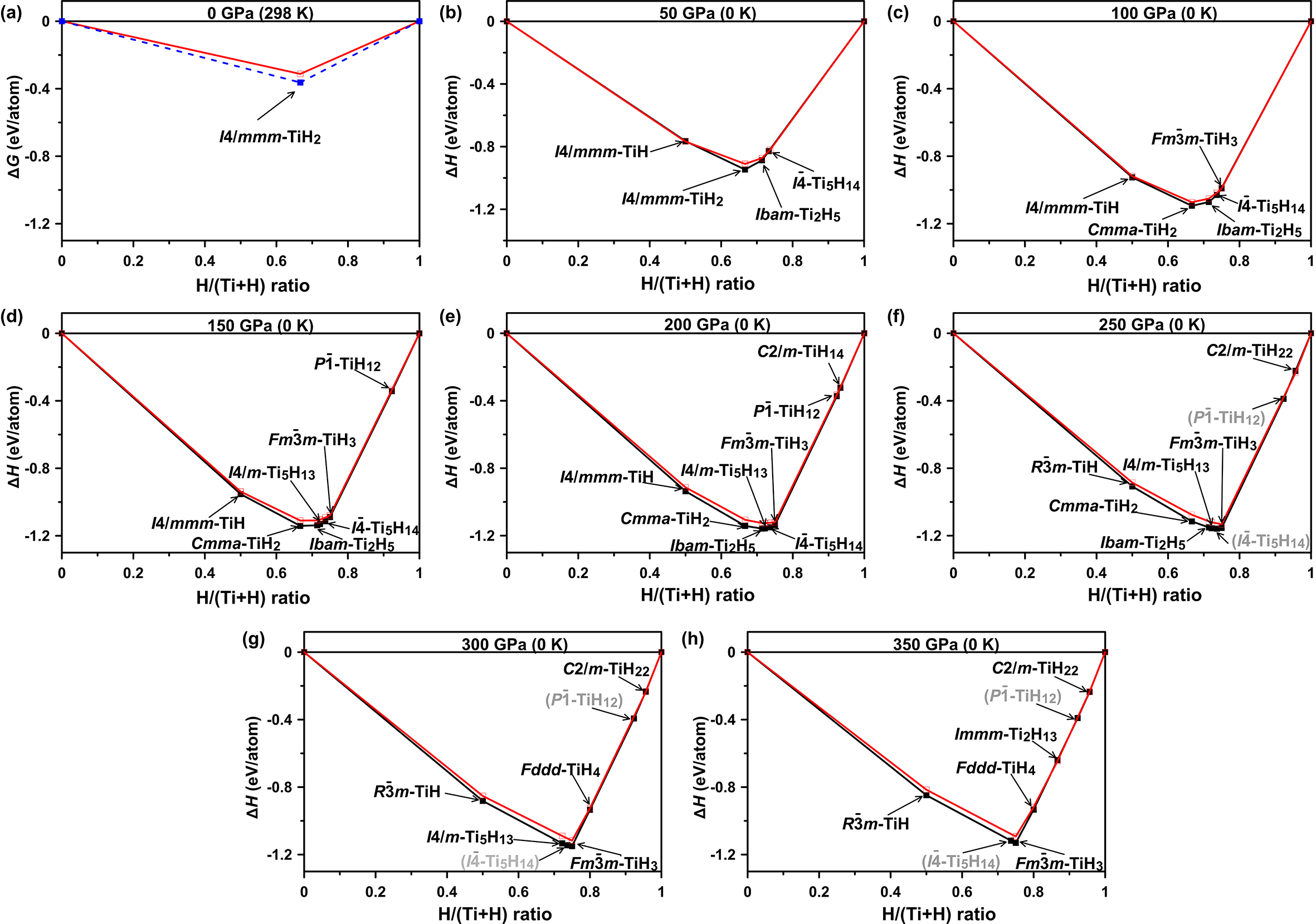}
\caption{\label{Fig1} (Color online) Convex-hull diagrams for the Ti--H system at (a) 0, (b) 50, (c) 100, (d) 150, (e) 200, (f) 250, (g) 300, and (h) 350 GPa. The blue dashed convex hull at 0 GPa shows the experimental result\cite{chase1998NIST}. Black lines with solid squares and red lines with open squares, respectively, represent the calculated results with and without including ZPE. The structures indicted in grey are those that lose stability after considering ZPE.}
\end{figure*}
\end{center}

Thermodynamic convex hulls for the Ti--H system at several pressures are shown in Fig.\ref{Fig1}. In our structure searching, experimentally reported \emph{I}4/\emph{mmm}-TiH$_{2}$ is found to be the only stable phase at zero pressure. Lattice parameters were optimized to be a=3.208 \AA{} and c=4.203 \AA{} at 0 GPa, which is in good accordance with the experimental data (a=3.163 \AA{} and c=4.391 \AA{})\cite{vennila2008phase}. The calculated Gibbs free energy of formation of \emph{I}4/\emph{mmm}-TiH$_{2}$ is -0.3123 eV/atom at zero pressure and 298 K [red convex hull in Fig.\ \ref{Fig1}(a)], which is in good agreement with the experimental value of -0.363 eV/atom \cite{chase1998NIST} [blue dashed convex hull in Fig.\ \ref{Fig1}(a)]. Besides \emph{I}4/\emph{mmm}-TiH and \emph{Fm}$\bar{3}$\emph{m}-TiH$_3$, which were already predicted by Zhuang\cite{zhuang2018effect} under high pressure, several other phases and stoichiometries, including \emph{R}$\bar{3}$\emph{m}-TiH, \emph{Cmma}-TiH$_2$, \emph{Ibam}-Ti$_2$H$_5$, \emph{I}4/\emph{m}-Ti$_5$H$_{13}$, \emph{I}$\bar{4}$-Ti$_5$H$_{14}$, \emph{Fddd}-TiH$_4$, \emph{Immm}-Ti$_2$H$_{13}$, \emph{P}$\bar{1}$-TiH$_{12}$, \emph{C}2/\emph{m}-TiH$_{14}$ and \emph{C}2/\emph{m}-TiH$_{22}$, are predicated at pressures up to 350 GPa. No subhydrides (Ti$_{x}$H$_{y}$, $x > y$) show up in the Ti--H system at any pressure. The enthalpies of formation with and without including zero-point energy (ZPE) are depicted by red lines with open squares and black lines with solid squares in Fig.\ref{Fig1}(b)--(h), respectively. Taking ZPE into account did not significantly change the basic shape of convex hulls, but did introduce some changes of the data at pressures above 250 GPa: considering ZPE made \emph{I}$\bar{4}$-Ti$_5$H$_{14}$ and \emph{P}$\bar{1}$-TiH$_{12}$ metastable [indicated in grey in Fig.\ \ref{Fig1}(f)--(h)] instead of stable structures above 250 GPa.

\begin{center}
\begin{figure}
\includegraphics[angle=0,width=1\linewidth]{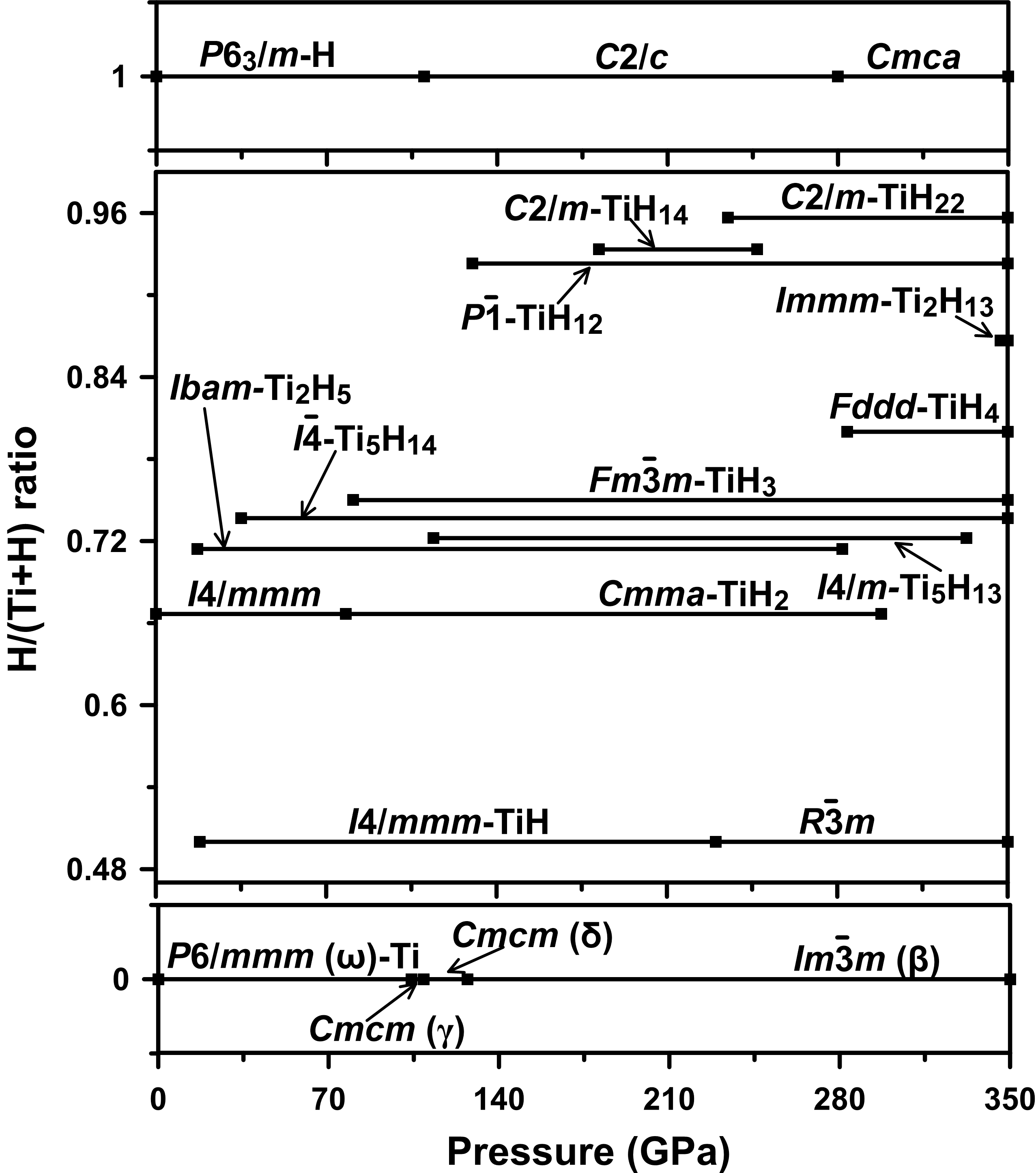}
\caption{\label{Fig2} (Color online) Pressure--composition phase diagram of the Ti--H system at pressure up to 350 GPa.}
\end{figure}
\end{center}

The pressure--composition phase diagram of Ti--H system is depicted in Fig.\ \ref{Fig2}. Based on our calculations, the phase transition sequence of Ti under high pressure is $\emph{P}6/\emph{mmm} (\omega) \xrightarrow{104\ \text{GPa}} \emph{Cmcm} (\gamma) \xrightarrow{109\ \text{GPa}} \emph{Cmcm} (\delta) \xrightarrow{127\ \text{GPa}}\mathit{Im}\bar{3}\mathit{m}(\beta)$. Although both $\gamma$-Ti and $\delta$-Ti have the same space group and contain 4 titanium atoms in their unit cell, the structure of $\gamma$-Ti is a distortion of $\omega$-Ti (hcp), while $\delta$-Ti, which is a body-centered one, is more similar to $\beta$-Ti (bcc). Under high pressure, \emph{P}6$_3$/\emph{m}-H transforms into \emph{C}2/\emph{c}-H at 110 GPa and further into \emph{Cmca}-H at 280 GPa. The crystal structures of these high-pressure structures are shown in Fig.\ \ref{Fig3}, and their structural parameters are listed in the Table S1 in Supplemental Material.

For TiH, it should be noted that the enthalpy of \emph{P}4$_2$/\emph{mmc}-TiH is lower than that of \emph{I}4/\emph{mmm}-TiH between 0-8 GPa [see also Fig.\ \ref{Fig4}] which is consistent with the aforementioned calculation\cite{zhuang2018effect}. However, \emph{P}4$_2$/\emph{mmc}-TiH is not thermodynamically stable from 0 to 8 GPa. With pressure increasing to 18 GPa, TiH (\emph{I}4/\emph{mmm}) begins to become stable and transforms into \emph{R}$\bar{3}$\emph{m}-TiH at 230 GPa. In addition, the calculations reveal a tetragonal (\emph{I}4/\emph{mmm}) to orthorhombic (\emph{Cmma}) phase transition in TiH$_2$ at 78 GPa. The high-pressure \emph{Cmma}-TiH$_2$ persists up to 298 GPa, above which TiH$_2$ is unstable. Note that enthalpy difference of \emph{Cmma}-TiH$_2$ and \emph{P}4/\emph{nmm}-TiH$_2$ is very small, due to the similarity of these two structures.

\begin{center}
\begin{figure*}
\includegraphics[angle=0,width=1\linewidth]{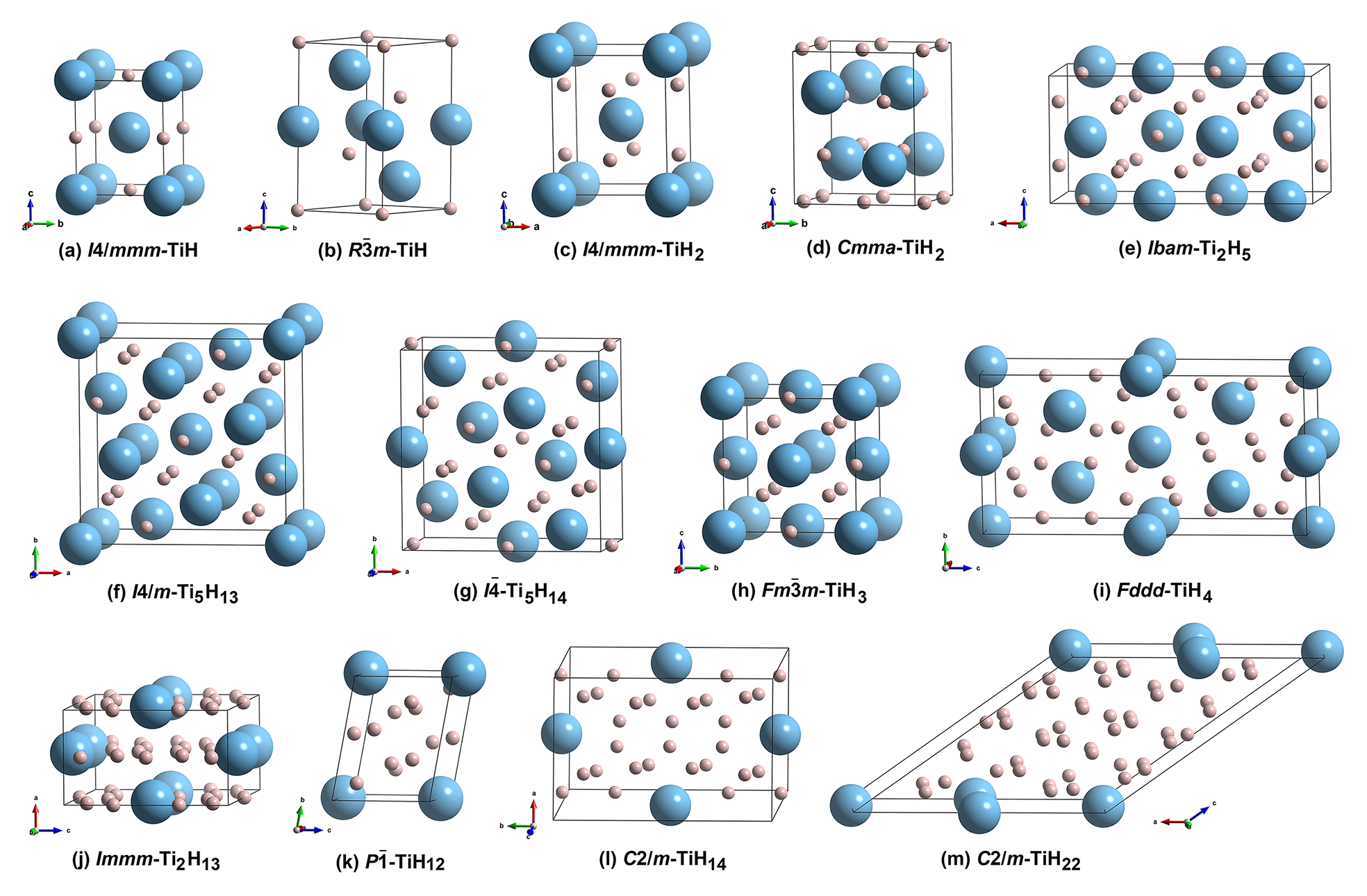}
\caption{\label{Fig3} (Color online) Crystal structures of titanium hydrides. Large spheres represent titanium atoms and small ones represent hydrogen atoms.}
\end{figure*}
\end{center}

\begin{center}
\begin{figure}
\includegraphics[angle=0,width=1\linewidth]{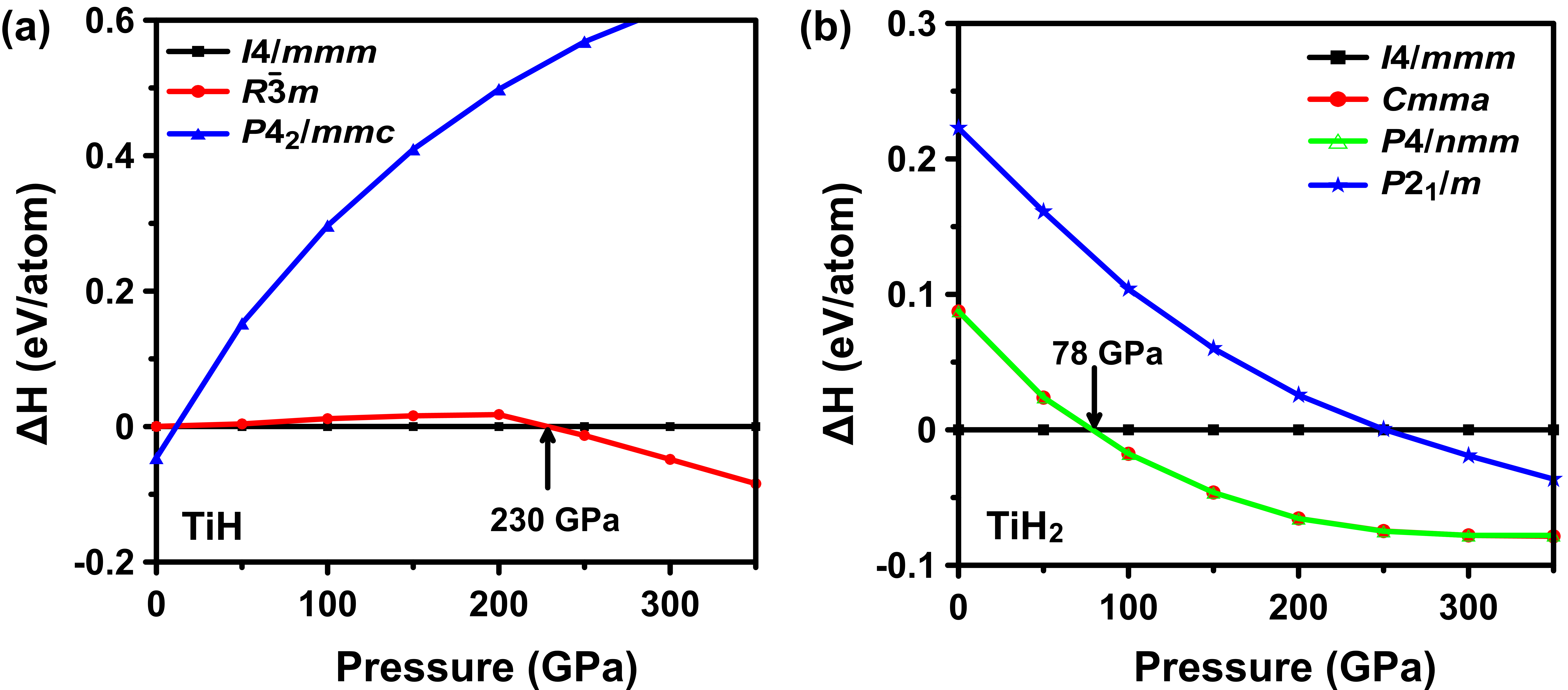}
\caption{\label{Fig4} (Color online) Enthalpy as a function of pressure (without ZPE) for phases of (a) TiH, referenced to the \emph{I}4/\emph{mmm}-TiH phase, and (b) TiH$_{2}$, referenced to the \emph{I}4/\emph{mmm}-TiH$_2$ one.}
\end{figure}
\end{center}

\begin{center}
\begin{figure*}
\includegraphics[angle=0,width=0.85\linewidth]{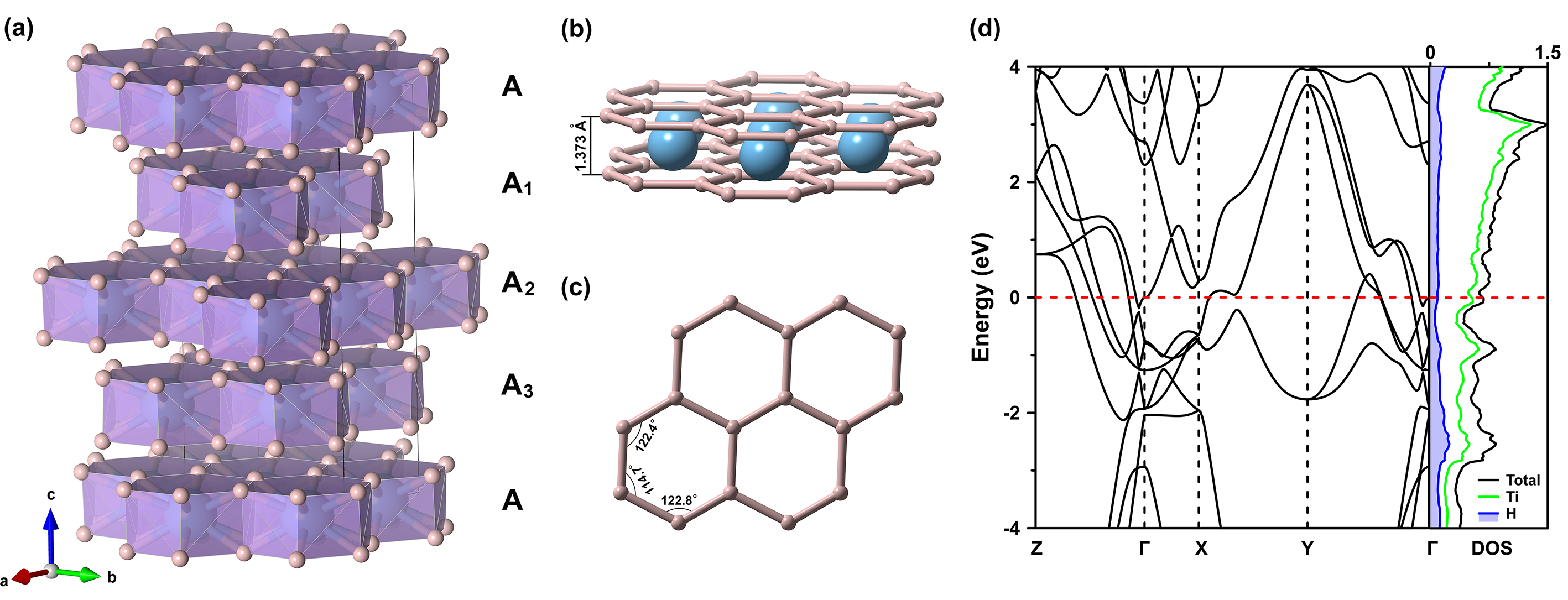}
\caption{\label{Fig5} (Color online) (a) Polyhedral representation of the \emph{Fddd}-TiH$_4$ structure. Ti-centered hexagonal prisms are shown in purple. (b) The fundamental ``sandwich" of TiH$_4$. (c) Distorted H-graphene layer in \emph{Fddd}-TiH$_4$, and (d) electronic band structure and density of states (DOS) of \emph{Fddd}-TiH$_4$ at 300 GPa; DOS is in the unit of states/formula/eV and Fermi energy (red dashed line) is set to zero.}
\end{figure*}
\end{center}

The structure of \emph{Fm}$\bar{3}$\emph{m}-TiH$_3$ has an fcc-sublattice of Ti atoms, all octahedral and tetrahedral voids of which are occupied by H atoms. It appears at 81 GPa, and continues to be stable up to at least 350 GPa. Another interesting structure is \emph{Fddd}-TiH$_4$ in which titanium atoms are sandwiched between two slightly distorted H-graphene layers [Fig.\ \ref{Fig5}(b)]. The structure of \emph{Fddd}-TiH$_4$ consists of such ``sandwiches'' in different orientations, forming AA$_1$A$_2$A$_3$AA$_1$A$_2$A$_3$A... stacking sequence [Fig.\ \ref{Fig5}(a)]. The distorted H-graphene layer is drawn in Fig.\ref{Fig5}(c) and the distance between two layers is 1.373 \AA{} at 350 GPa, as seen in Fig.\ \ref{Fig5}(b). \emph{Immm}-TiH$_6$, which is reported to be stabilized above 175 GPa\cite{zhuang2018effect}, is actually a metastable phase and decomposes into \emph{Fm}$\bar{3}$\emph{m}-TiH$_3$ and \emph{P}$\bar{1}$-TiH$_{12}$ at high pressure according to our results. Ti$_2$H$_{13}$, a stoichiometry close to TiH$_6$, emerges on the phase diagram at 347 GPa and adopts a \emph{Immm} structure. TiH$_{14}$ is stable from 182 to 247 GPa. Although both TiH$_{14}$ and SnH$_{14}$\cite{esfahani2016superconductivity} crystallize in $\emph{C}2/\emph{m}$ space group, their structures and their hydrogen sublattices are different.

\begin{center}
\begin{figure*}
\includegraphics[angle=0,width=0.85\linewidth]{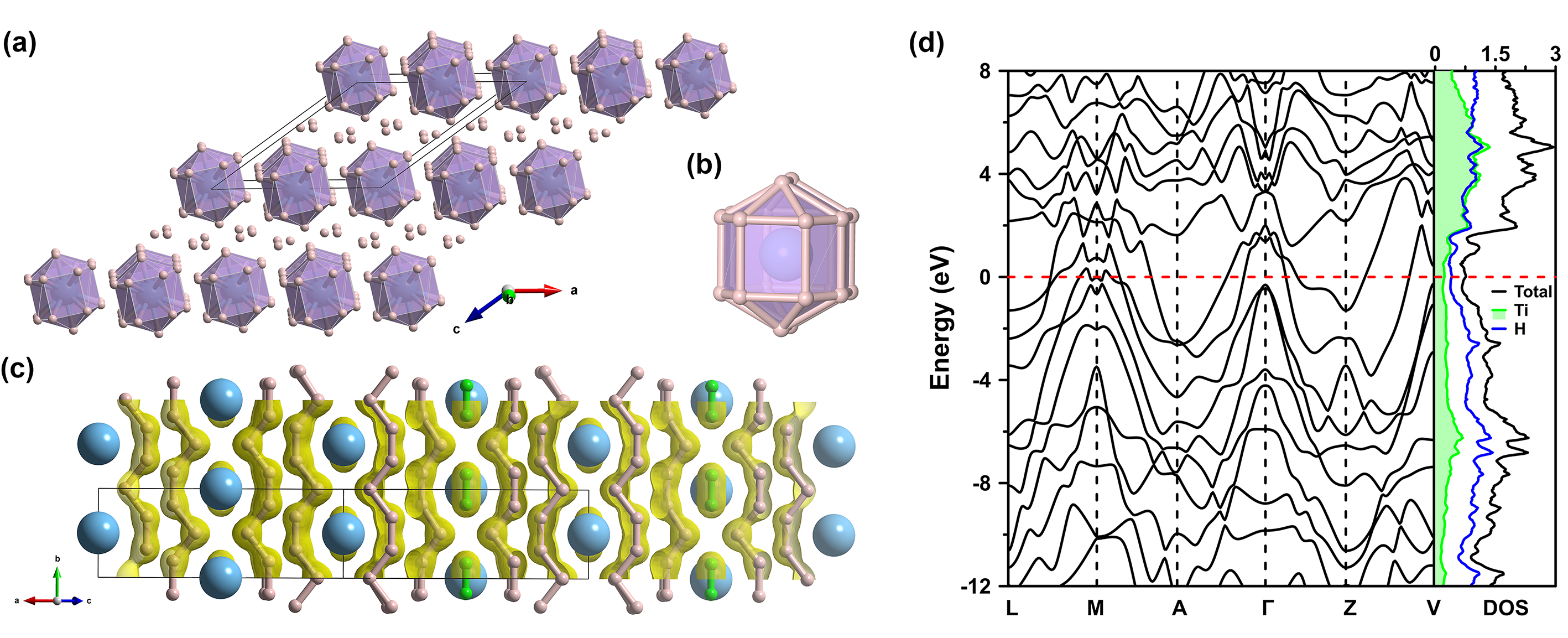}
\caption{\label{Fig6} (Color online) (a) Polyhedral representation of the \emph{C}2/\emph{m}-TiH$_{22}$ structure. (b) Expanded view of a H$_{20}$ cage encapsulating a Ti atom. (c) ELF isosurface (ELF=0.7) for \emph{C}2/\emph{m}-TiH$_{22}$ at 350 GPa. Green and pink atoms represent molecular hydrogen and armchair-like hydrogen chains, respectively. (d) Electronic band structure and DOS for \emph{C}2/\emph{m}-TiH$_{22}$ at 350 GPa. }
\end{figure*}
\end{center}

The most interesting part is that besides hydrogen-rich TiH$_{14}$ stoichiometry, an \textit{extremely} H-rich structure TiH$_{22}$ is identified to be thermodynamically stable in a monoclinic structure at pressures above 235 GPa. To the best of our knowledge, \emph{C}2/\emph{m}-TiH$_{22}$ presently is the second hydrogen-richest hydrides known or predicted to date, after metal hydride \emph{C}2/\emph{c}-YH$_{24}$\cite{peng2017hydrogen}. The polyhedral crystal structure representation of \emph{C}2/\emph{m}-TiH$_{22}$ [depicted in Fig.\ \ref{Fig6}(a)] exhibits alternations of H$_2$ molecules and TiH$_{20}$ polyhedra. Titanium is encapsulated in H$_{20}$ cages with Ti-H distances are 1.62--1.66 \AA{} at 350 GPa, as shown in Fig.\ \ref{Fig6}(b). The band structure and density of states (DOS) of \emph{C}2/\emph{m}-TiH$_{22}$ at 350 GPa [Fig.\ \ref{Fig6}(d)] indicate metallicity of TiH$_{22}$. The total DOS of \emph{C}2/\emph{m}-TiH$_{22}$ near the Fermi level \emph{N}($\varepsilon_{F}$) mostly comes from H atoms, which is opposite to \emph{Fddd}-TiH$_4$ [Fig.\ \ref{Fig5}(d)]. The coexistence of molecular hydrogen with an H--H distance of 0.819 \AA{} and armchair-like hydrogen chain is clearly revealed by the electron localization function (ELF). As shown in ELF Fig.\ \ref{Fig6}(c), the regions with ELF values of 0.7 include H$_2$ molecules and armchair-like hydrogen chain, which indicates strong covalent bonding between hydrogen atoms.

\begin{center}
\begin{figure*}
\includegraphics[angle=0,width=1\linewidth]{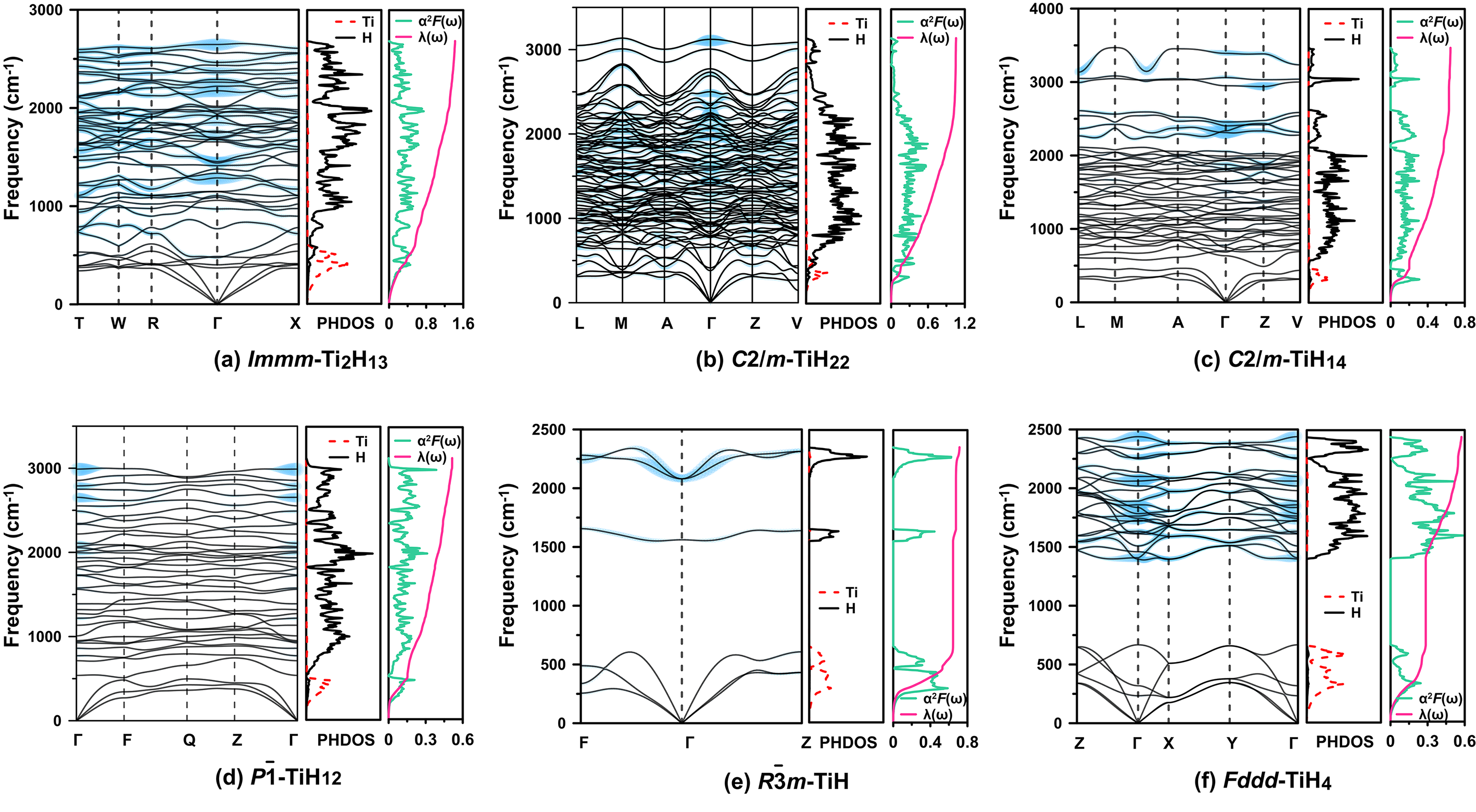}
\caption{\label{Fig7} (Color online) Phonon dispersion curves, phonon density of states projected onto selected atoms, Eliashberg spectral function $\alpha^{2}\emph{F}(\omega)$, and the electron--phonon coupling (EPC) parameter $\lambda$ for (a) \emph{Immm}-Ti$_2$H$_{13}$ at 350 GPa, (b) \emph{C}2/\emph{m}-TiH$_{22}$ at 250 GPa, (c) \emph{C}2/\emph{m}-TiH$_{14}$ at 200 GPa, (d) \emph{P}$\bar{1}$-TiH$_{12}$ at 350 GPa, (e) \emph{R}$\bar{3}$\emph{m}-TiH at 200 GPa and (f) \emph{Fddd}-TiH$_4$ at 350 GPa. The magnitude of the phonon linewidths is indicated by the size of the blue open circles with the radius proportional to the respective coupling strength.}
\end{figure*}
\end{center}

Based on the calculated phonon dispersion spectrum at high pressure [displayed in Fig.\ref{Fig7} and Fig. S1], no imaginary vibrational frequencies are found in the whole Brillouin zone, indicating the dynamical stability of all the predicted structures. Phonon dispersion curves, phonon density of states, phonon linewidths $\gamma _{\mathbf{q}\nu}$, Eliashberg phonon spectral function $\alpha ^{2}F(\omega )$, and the electron-phonon coupling parameter $\lambda$ of \emph{Immm}-Ti$_2$H$_{13}$ , \emph{C}2/\emph{m}-TiH$_{22}$ , \emph{I}$\bar{4}$-Ti$_5$H$_{14}$, \emph{P}$\bar{1}$-TiH$_{12}$, \emph{R}$\bar{3}$\emph{m}-TiH and \emph{Fddd}-TiH$_4$, at selected pressures are depicted in Fig.\ref{Fig7}. As expected (due to atomic masses), low-frequency modes are mostly related to Ti atoms whereas high-frequency modes are dominated by vibrations of H ones. Moreover, in all of these six structures, the $\gamma _{\mathbf{q}\nu}$ of branches near the $\Gamma$ point are much greater than those elsewhere in the Brillouin zone. The total $\lambda$ of \emph{R}$\bar{3}$\emph{m}-TiH is mainly contributed by the acoustic modes, whereas those of the other five structures are dominated by optical branches.

\begin{center}
\begin{figure*}
\includegraphics[angle=0,width=1\linewidth]{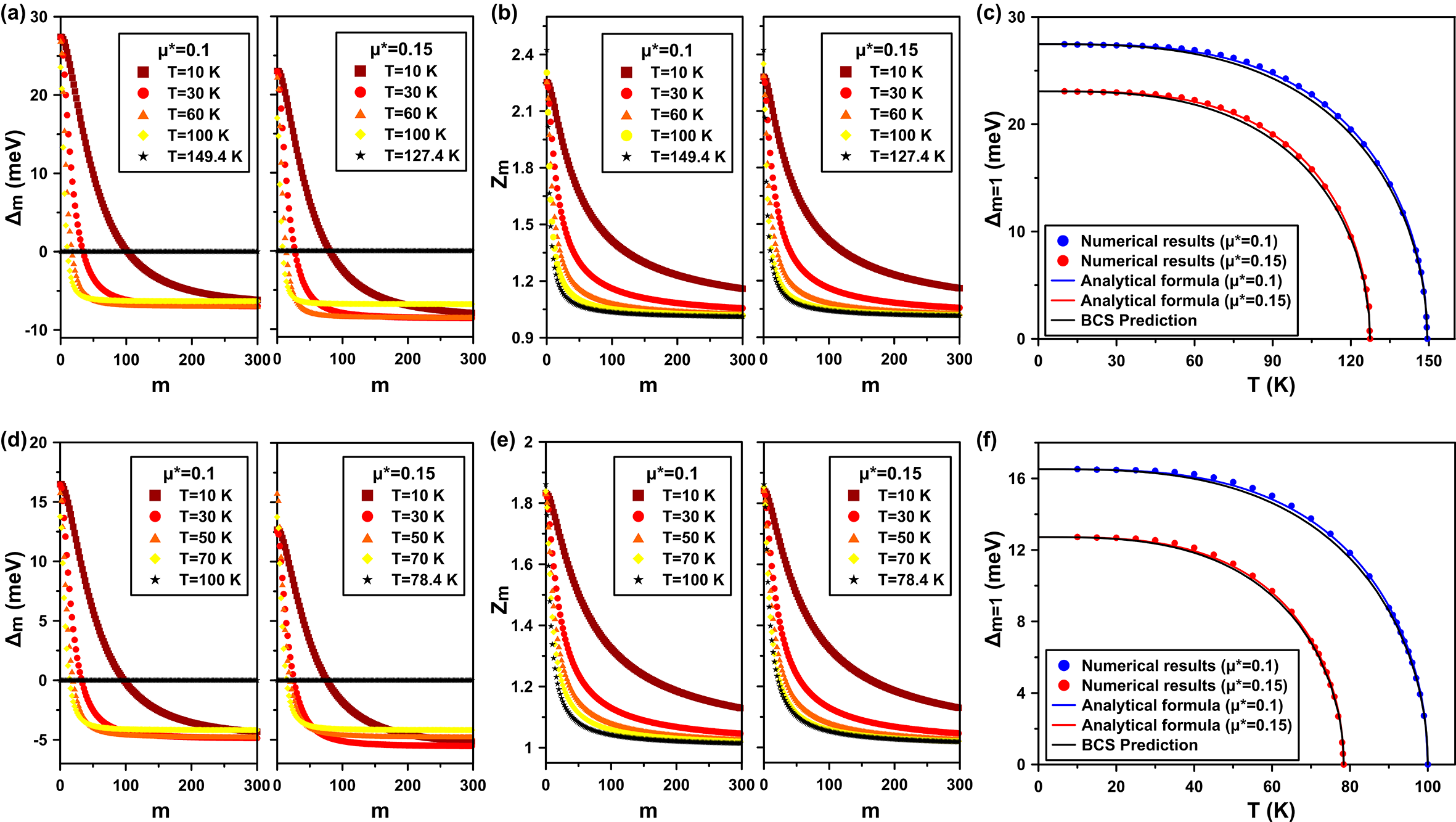}
\caption{\label{Fig8} (Color online) Dependence of the superconducting order parameter on the number m for select values of temperature and Coulomb pseudopotential for (a) \emph{Immm}-Ti$_2$H$_{13}$ at 350 GPa and (d) \emph{C}2/\emph{m}-TiH$_{22}$ at 250 GPa. The wave function renormalization factor Z$_\text{m}$ on the imaginary axis for select values of temperature and Coulomb pseudopotential for (b) \emph{Immm}-Ti$_2$H$_{13}$ and (e) \emph{C}2/\emph{m}-TiH$_{22}$. The influence of temperature on the maximum value of the order parameter ($\Delta_{m=1}$) for selected $\mu$$^{*}$ of (c) \emph{Immm}-Ti$_2$H$_{13}$ and (f) \emph{C}2/\emph{m}-TiH$_{22}$. Solid circles correspond to the exact numerical solutions of the Eliashberg equations, the red and blue lines represent the results obtained using analytical formulas [Eq.\ref{eq9}]. Black lines are predicted by BCS model [Eq.\ref{eq9}, where $\Gamma$=3] }
\end{figure*}
\end{center}

We further probe superconductivity of these hydrides, using BCS theory. The calculated superconducting properties are summarized in Table \ref{Tab1}. All of the predicted titanium hydrides exhibit superconductivity at high pressures. The highest \emph{T}$_{c}$ of titanium hydrides are possessed by \emph{Immm}-Ti$_2$H$_{13}$, \emph{C}2/\emph{m}-TiH$_{22}$, \emph{I}$\bar{4}$-Ti$_5$H$_{14}$, \emph{P}$\bar{1}$-TiH$_{12}$, \emph{R}$\bar{3}$\emph{m}-TiH and \emph{Fddd}-TiH$_4$. \emph{I}4/\emph{mmm}-TiH$_2$ exhibits low \emph{T}$_{c}$ values (3 mK, $\mu^{*}$=0.1) at 50 GPa. On the other hand, superconductivity of titanium monohydride (TiH) comes largely from strong coupling of the electrons with Ti vibrations, and coupling with H vibrations becomes more important as H content increases. Intriguingly, it is \emph{Immm}-Ti$_2$H$_{13}$ instead of \emph{C}2/\emph{m}-TiH$_{22}$ that possesses the highest \emph{T}$_{c}$ among searched titanium hydrides. The results from the previous studies\cite{liu2017potential,zhong2016tellurium} suggest higher hydrogen content in the binary hydrides is one of the necessary prerequisites to obtain higher \emph{T}$_{c}$ value. This is not necessarily always the case; the hydrogen content in \emph{C}2/\emph{m}-TiH$_{22}$ is much higher than in \emph{Immm}-Ti$_2$H$_{13}$. Indeed, $\omega_{\textit{log}}$ of \emph{C}2/\emph{m}-TiH$_{22}$ is larger than that of \emph{Immm}-Ti$_2$H$_{13}$. However, this is offset by the lower $\lambda$ of \emph{C}2/\emph{m}-TiH$_{22}$ ($\lambda$=0.861) compared with that of \emph{Immm}-Ti$_2$H$_{13}$ ($\lambda$=1.423). The $\alpha ^{2}F(\omega)$ was used for numerically solving the Eliashberg equations and the obtained \emph{T}$_{c}$ of \emph{Immm}-Ti$_2$H$_{13}$ is in the range 110.4--131.2 K ($\lambda$=1.423, $\mu^{*}$=0.1--0.15) at 350 GPa.

\begin{table*}
\centering
\caption{The EPC parameter $\lambda$, electronic density of states at Fermi level \emph{N}($\varepsilon_{F}$) (states/Ry/cell), the logarithmic average phonon frequency $\omega_{log}$ (K) and superconducting critical temperatures \emph{T}$_{c}$ (K) for titanium hydrides at different pressure \emph{P} (GPa). \emph{T}$_{c}$ values are given for $\mu^{*}$=0.1 and \emph{T}$_{c}$ in brackets are for $\mu$=0.15; \emph{T}$_{c}$ (McM) is the numerical solution of solving the imaginary axis Eliashberg equation, T$_{c}$ (A--D) is calculated from the Allen--Dynes equation and \emph{T}$_{c}$ (E) is obtained by Allen--Dynes modified McMillan equation.}
\begin{tabular}{ c c c c c c c c  }   
\hline \hline
Compound                            & \emph{P}  &$\lambda$ & \emph{N}($\varepsilon_{F}$)  & $\omega_{log}$ & \emph{T}$_{c}$ (McM)    & \emph{T}$_{c}$ (A--D)       & \emph{T}$_{c}$ (E)  \\
\hline
\emph{C}2/\emph{m}-TiH$_{22}$       & 350 & 0.861    &  4.765  &  1677.4  & 90.7 (65.0)   & 93.6 (67.3)   &  100.0 (78.4)        \\
                                    & 250 & 1.057    &  4.867  &  1296.2  & 98.1 (76.7)   & 103.1 (80.7)  &  110.2 (91.3)       \\
\emph{C}2/\emph{m}-TiH$_{14}$       & 200 & 0.645    &   5.243 &  1201.7  & 33.9 (19.7)   & 35.0 (20.3)   &  35.9 (25.0)         \\
\emph{P}$\bar{1}$-TiH$_{12}$        & 350 & 0.514    &   3.213 &  1357.6  & 18.4 (7.8)    &  18.8 (8.0)   &  19.5 (11.5)         \\
                                    & 150 & 0.403    &   3.748 &  1074.8  & 4.7 (1.0)     &  4.8 (1.0)    &  5.4 (2.4)        \\
\emph{Immm}-Ti$_2$H$_{13}$          & 350 & 1.423    &   10.028&  1101.3  & 119.3 (100.8) &  131.2 (110.4)& 149.4 (127.4)          \\
\emph{Fddd}-TiH$_4$                 & 350 & 0.574    &   3.803 &  1034.4  & 20.6 (10.4)   &  21.2 (10.7)  &  20.1 (6.2)       \\
\emph{Fm}$\bar{3}$\emph{m}-TiH$_3$  & 100 & 0.528    &   6.798 &  459.4   & 6.9 (3.1)     &  7.1 (3.1)    & 7.5 (4.7)      \\
\emph{I}$\bar{4}$-Ti$_5$H$_{14}$    & 350 & 0.411    &  20.738 &  479.9   & 2.4 (0.6)     &  2.4 (0.6)    &  2.8 (1.4)       \\
                                    & 50  & 0.477    &  35.785 &  525.6   & 5.3 (1.9)     &  5.4 (2.0)    &  5.8 (3.3)     \\
\emph{I}4/\emph{m}-Ti$_5$H$_{13}$   & 300 & 0.470    &  22.211 &  412.6   & 3.9 (1.4)     &  4.0 (1.4)    &  4.4 (2.7)      \\
                                    & 150 & 0.406    &  27.253 &  450.8   & 2.1 (0.5)     &  2.1 (0.5)    &  2.5 (1.1)    \\
\emph{Ibam}-Ti$_2$H$_5$             & 250 & 0.504    &  9.910  &  363.3   & 4.6 (1.9)     &  4.7 (1.9)    &  5.1 (3.2)      \\
                                    & 50  & 0.564    &  12.852 &  365.0   & 6.9 (3.4)     &  7.1 (3.5)    &  6.9 (4.6)     \\
\emph{Cmma}-TiH$_2$                 & 250 & 0.509    &  3.604  &  434.3   & 5.7 (2.4)     &  5.8 (2.4)    &  6.0 (3.7)     \\
\emph{I}4/\emph{mmm}-TiH$_2$        & 50  & 0.227    &  3.687  &  0.0 (0.0)   & 0.0 (0.0) &  0.0 (0.0)    &  0.0 (0.0)      \\
\emph{R}$\bar{3}$\emph{m}-TiH       & 350 & 0.714    &  4.328  &  597.3   & 21.8 (13.9)   &  22.7 (14.4)  & 23.9 (18.2)       \\
\emph{I}4/\emph{mmm}-TiH            & 200 & 0.991    &  6.303  &  264.3   & 18.1 (13.8)   &  19.5 (14.8)  & 22.5 (18.9)       \\
                                    & 50  & 1.013    &  8.716  &  71.0    & 5.0 (3.9)     &    5.4 (4.1)  & 12.6 (10.0)        \\
\hline\hline
\end{tabular}
\label{Tab1}
\end{table*}

For \emph{Immm}-Ti$_2$H$_{13}$ at 350 GPa and \emph{C}2/\emph{m}-TiH$_{22}$ at 250 GPa, the dependence of the maximum value of the order parameter on temperature for selected $\mu^{*}$ is presented in Fig.\ \ref{Fig8}(c) and (f). The maximum value of order parameter $\Delta_{m=1}$ decreases with the growth of \emph{T} and $\mu^{*}$. On the basis of these results, $\Delta_{m=1}$ value can be characterized analytically by means of the phenomenological formula
\begin{equation}\label{eq9}
{\textstyle \Delta_{m=1}(T,\mu^{*})=\Delta _{m=1}(T_{0},\mu^{*})\sqrt{1-\left ( \frac{T}{T_{c}} \right )^{\Gamma }}}.
\end{equation}
For the maximum value of order parameter $\Delta_{m=1}$, we obtained the estimation of temperature exponent for \emph{Immm}-Ti$_2$H$_{13}$ ($\Gamma$=3.25 for $\mu^{*}$=0.1;  $\Gamma$=3.31 for $\mu^{*}$=0.15) and \emph{C}2/\emph{m}-TiH$_{22}$ ($\Gamma$=3.21 for $\mu^{*}$=0.1;  $\Gamma$=3.16 for $\mu^{*}$=0.15). It is clear that the temperature dependence of maximum order parameter obtained in Eliashberg equations only differs a little bit from the results estimated by the BCS theory, where $\Gamma$=3 (Ref.\ \onlinecite{eschrig2001theory}). It can be seen that the values of order parameter strongly decrease together with the increase of the Coulomb pseudopotential [Fig.\ \ref{Fig8}(a) and (d)]. On the other hand, the influence of the Coulomb pseudopotential $\mu^{*}$ on the wavefunction renormalization factor [Fig.\ \ref{Fig8}(b) and (e)] is significantly weaker. Through comparison among above three approaches of calculating \emph{T}$_{c}$, it can be seen that two analytical results generally underestimate \emph{T}$_{c}$, especially for the high values of the Coulomb pseudopotential. Moreover, the Allen--Dynes formula much better reproduces the numerical results than the modified McMillan expression. Note that anharmonicity, which usually decreases \emph{T}$_{c}$, is not included in our calculations.

The influence of pressure on \emph{T}$_{c}$ has been widely discussed before. Theoretical studies of some systems\cite{yu2015pressure,liu2017potential,kvashnin2018high} show that \emph{T}$_{c}$ will decrease with increasing pressure; some\cite{kim2010general,li2014metallization} report \emph{T}$_{c}$ to increase with pressure; and others\cite{zhang2015phase,zheng2018structural} reveal negligible pressure dependence. The first two situations can be reflected in Ti--H system. For example, the \emph{T}$_{c}$ of \emph{C}2/\emph{m}-TiH$_{22}$, \emph{I}$\bar{4}$-Ti$_5$H$_{14}$, \emph{Ibam}-Ti$_2$H$_5$ and \emph{I}4/\emph{mmm}-TiH decrease with pressure, whereas \emph{T}$_{c}$ of \emph{P}$\bar{1}$-TiH$_{12}$ and \emph{I}4/\emph{m}-Ti$_5$H$_{13}$ increase. One of the important factors explaining the effect of pressure on \emph{T}$_{c}$ is related to phonon softening. In case of \emph{C}2/\emph{m}-TiH$_{22}$, phonon modes around the A and Z points harden with pressure [see Figs.\ \ref{Fig7}(b) and S1(a)]. The same tendency can also be seen around the P and N points in the Brillouin zone of \emph{I}$\bar{4}$-Ti$_5$H$_{14}$ on increasing pressure. For \emph{P}$\bar{1}$-TiH$_{12}$, phonon modes around the $\Gamma$ and Z point soften with pressure. This means that phonon modes around high-symmetry points harden with pressure, leading to a decrease of the value of \emph{T}$_{c}$. On the contrary, phonon softening with pressure gives rise to the increase of \emph{T}$_{c}$.

\section{Conclusions}
In order to discover high-\emph{T}$_{c}$ superconductors, the Ti--H system at pressures up to 350 GPa was systematically explored using the \textit{ab initio} evolutionary algorithm USPEX. A phase (\emph{R}$\bar{3}$\emph{m}-TiH) and several stoichiometries (\emph{C}2/\emph{m}-TiH$_{22}$, \emph{P}$\bar{1}$-TiH$_{12}$, \emph{Immm}-Ti$_2$H$_{13}$, \emph{Fddd}-TiH$_4$, \emph{I}$\bar{4}$-Ti$_5$H$_{14}$, \emph{I}4/\emph{m}-Ti$_5$H$_{13}$ and \emph{Ibam}-Ti$_2$H$_5$) were predicted, and found to be dynamically stable in their predicted pressure ranges of stability. With increasing pressure, \emph{I}4/\emph{mmm}-TiH transforms into \emph{R}$\bar{3}$\emph{m}-TiH at 230 GPa, and \emph{I}4/\emph{mmm}-TiH$_2$ into \emph{Cmma}-TiH$_2$ at 78 GPa. \emph{Cmma}-TiH$_2$ is structurally similar to \emph{P}4/\emph{nmm}-TiH$_2$. \emph{C}2/\emph{m}-TiH$_{22}$ has the highest hydrogen content among all titanium hydrides, and contains TiH$_{20}$ cages. The estimated \emph{T}$_{c}$ of \emph{Immm}-Ti$_2$H$_{13}$ is 127.4--149.4 K ($\mu^{*}$=0.1--0.15) at 350 GPa, which is actually higher than \emph{T}$_{c}$ of the aforementioned \emph{C}2/\emph{m}-TiH$_{22}$ of 91.3--110.2 K at 250 GPa. Superconductivity of \emph{Immm}-Ti$_2$H$_{13}$ mainly arises from both strong coupling of the electrons with H vibrations and the large logarithmic average phonon frequency. The accuracy of three methods for estimating the \emph{T}$_{c}$ were compared. Taking solution of the Eliashberg equations as standard, the estimated \emph{T}$_{c}$ from Allen--Dynes formula is more accurate than that from the modified McMillan expression. The constructed pressure--composition phase diagram and the analysis of superconductivity of titanium hydrides will motivate future experimental synthesis of titanium hydrides and studies of their high-temperature superconductivity.

\section{acknowledgments}

J.\ M.\ M.\ acknowledges startup support from Washington State University and the Department of Physics and Astronomy thereat. A.\ R.\ O.\ thanks Russian Science Foundation (grant 19-72-30043) for financial support.

\end{document}